\documentclass[a4paper,conference]{IEEEtran}
\IEEEoverridecommandlockouts
\usepackage[left=1.62cm,right=1.62cm,top=1.9cm]{geometry}
\usepackage{cite} 
\usepackage{amsmath,amssymb,amsfonts,balance}
\usepackage{algorithmic}
\usepackage{graphicx}
\usepackage{textcomp}
\usepackage{xcolor}
\def\BibTeX{{\rm B\kern-.05em{\sc i\kern-.025em b}\kern-.08em
    T\kern-.1667em\lower.7ex\hbox{E}\kern-.125emX}}
\begin{document}

\title{Markov Chain-based Model of \\ Blockchain Radio Access Networks 
\thanks{This work has received funding from the Smart Networks and Services Joint Undertaking (SNS JU) under the European Union’s Horizon Europe research and innovation programme under Grant Agreement No. 101096456 (NANCY).}
}

\author{\IEEEauthorblockN{Vasileios Kouvakis\IEEEauthorrefmark{1}, Stylianos E. Trevlakis\IEEEauthorrefmark{1}, Alexandros-Apostolos A. Boulogeorgos\IEEEauthorrefmark{2}, Hongwu Liu\IEEEauthorrefmark{3}, \\ Theodoros A. Tsiftsis\IEEEauthorrefmark{4}\IEEEauthorrefmark{1}, and Octavia A. Dobre\IEEEauthorrefmark{5}
 } 
\vspace{0.35mm}
\IEEEauthorblockA{\IEEEauthorrefmark{1}Department of Research and Development, InnoCube P.C., 17th Noemvriou 79, Thessaloniki 55535, Greece.\\
\IEEEauthorrefmark{2}Department of Electrical and Computer Engineering, University of Western Macedonia, Kozani 50100, Greece.\\
\IEEEauthorrefmark{3} School of Information Science and Electrical Engineering, Shandong Jiaotong University, Jinan 250357, China.\\
\IEEEauthorrefmark{4}Department of Informatics \& Telecommunications, University of Thessaly, Lamia 35100, Greece.\\
\IEEEauthorrefmark{5} Faculty of Engineering and Applied Science, Memorial University, St. John’s, NL A1B 3X9, Canada.\\
\vspace{0.35mm}
\IEEEauthorblockA{Emails: \{kouvakis, trevlakis\}@innocube.org, al.boulogeorgos@ieee.org, liuhongwu@sdjtu.edu.cn, \\ tsiftsis@uth.gr, odobre@mun.ca
}}}

\maketitle

\begin{abstract}
Security has always been a priority, for researchers, service providers and network operators when it comes to radio access networks (RAN). One wireless access approach that has captured attention is blockchain enabled RAN (B-RAN) due to its secure nature. This research introduces a framework that integrates blockchain technology into RAN while also addressing the limitations of state-of-the-art models. The proposed framework utilizes queuing and Markov chain theory to model the aspects of B-RAN. An extensive evaluation of the models performance is provided, including an analysis of timing factors and a focused assessment of its security aspects. The results demonstrate reduced latency and comparable security making the presented framework suitable for diverse application scenarios.
\end{abstract}

\begin{IEEEkeywords}
Alternative history attack, B-RAN, blockchain, Markov chain, queuing theory, radio access networks, security, timing.
\end{IEEEkeywords}

\section{Introduction}
In the changing world of telecommunications, radio access networks (RAN) play a pivotal role in facilitating smooth wireless communication linking countless devices. As RAN technology progresses it becomes ever more important to have security and privacy measures, in place to protect against emerging risks and vulnerabilities~\cite{caso2024experimentation,mesodiakaki20236g,Trevlakis2023}. Security, in the RAN has always been a focus of interest for network operators, service providers and researchers alike. The potential risks of access, data breaches and disruptions to the network can have implications for the confidentiality and integrity of communications within the RAN~\cite{Cao2021}. In this context the integration of technology emerges as a solution to address these security concerns. Originally designed for cryptocurrencies blockchain has proven its effectiveness, in offering decentralized tamper resistant solutions. By being decentralized blockchain ensures that no single entity controls the network thereby reducing the risks associated with points of failure and unauthorized access.
 
The use of Blockchain in RAN has been a topic of interest in the field of telecommunications security \cite{Dulaimi2023}. Many studies have explored how blockchain technology can help address security challenges in network domains making it an important area of research within the broader telecommunications landscape. In~\cite{Wang2021}, the authors thoroughly investigate the integration of blockchain into communications and propose a secure B-RAN framework for 6G networking. They also introduce a framework, for analyzing block structured Markov processes, which extends existing models to include phase type service times and transaction arrivals. Markov Chain (MC) models are commonly used by researchers to study B-RAN systems as seen in~\cite{Ling2019, Ling2021}. These studies introduce a B-RAN architecture, describe its workflow, and establish a queuing theory-based MC model to characterize system latency and security in B-RANs.

Furthermore, a common focus on estimating the most suitable block size has been identified in various recent works~\cite{Li2018, Wilhelmi2022, YixinLi2020, Wilhelmi2021}. Specifically, \cite{Li2018} concentrates on depicting mining processes and stages of construction, while~\cite{Wilhelmi2022} describes timers and forks within its latency model. In order to overcome challenges related to blockchain forking in generation wireless networks, the authors of~\cite{YixinLi2020} introduce a proposal for block access control. This approach efficiently manages block transmission enhances transaction throughput and imposes limitations on requirements. Using a MC model this study evaluates the performance of a network with block access control to demonstrate its effectiveness and highlight any limitations it may have. Similarly, in~\cite{Wilhelmi2021}, a MC based model of blockchain is employed in order to minimize the impact of latency on system stability through batch service queueing. This study advocates for the use of the Bianchi model in evaluating service latency with the aim of gaining insights, into blockchain network performance and reliability.

\begin{figure*}
    \centering
    \includegraphics[width=0.9\linewidth]{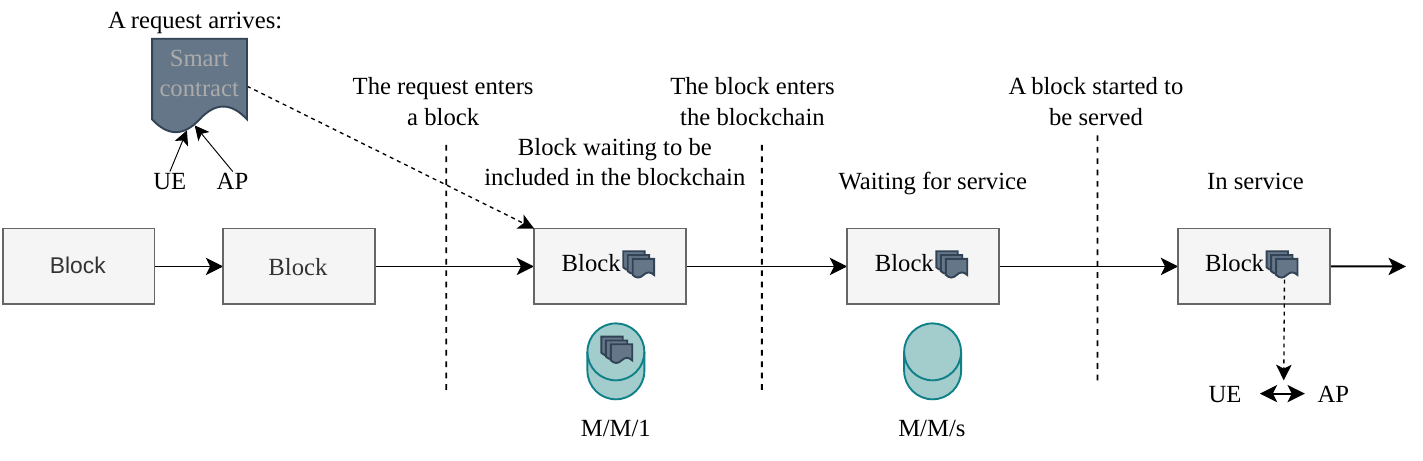}
    \caption{Procedural illustration of servicing a UE in B-RAN.}
    \label{Fig:BRAN}
\end{figure*}
This paper explores the different aspects of B-RAN, investigating how this innovative technology can greatly enhance security and privacy measures. In order to improve the modeling of B-RAN, a departure from the conventional approaches is proposed. The proposed framework involves creating two queuing models and a MC model. By implementing a mechanism to incorporate the number of transactions in a block as well as the possibility of rejected transactions, it enhances the accuracy of conventional B-RAN modeling. As a result, the timing performance of the model is greatly enhanced while also maintaining security and privacy in the same levels. This expansion of conventional B-RAN models aims to examine and support various scenarios; thus, enhancing its versatility and ability to address a wider range of potential applications within RAN.

The remainder of the paper is structured as follows. Section~\ref{S:system_model} serves as an introduction to the architectural and operational mechanisms of the blockchain, encompassing queuing models and the MC model. In Section~\ref{S:performance}, a performance evaluation of the model is presented, featuring a detailed examination of temporal aspects and a dedicated exploration of the applied attack model. Section~\ref{S:results} illustrates the numerical outcomes derived from the simulation of the proposed model, coupled with interesting conversations. Finally, Section~\ref{S:conclusions} encapsulates the conclusions, synthesizing the information gleaned from the aforementioned data and results.

\section{Markov chain-based model}\label{S:system_model}
The considered system model, which is presented in~Fig.~\ref{Fig:BRAN}, models the B-RAN by using two queues. The first queue handles incoming requests that are waiting to be included into the next blockchain block, while the second manages the mined blocks that contain confirmed requests, which wait to be serviced. In more detail, the first queue is structured according to the principles of an M/M/1 queue. In this model, the arrival of requests follows a Poisson distribution, with rate $\lambda_a$, while the processing times of these requests are governed by memoryless exponential distributions with rate $\lambda_b$. It should be noted that each individual request is processed within a discrete block and the system is designed to handle a maximum of $k$ requests in a single block. This specification underscores a finite capacity constraint, setting distinct parameters for the operational dynamics of the system. Moreover, the second queue is modeled as an M/M/s queue, where $s$ denotes the maximum number of access links. Similar to the M/M/1 model, requests enter this queue following a Poisson distribution, and their processing times are characterized by memoryless exponential distributions. 

Within the context of B-RAN, the comprehensive system configuration can be modeled through queuing theory, which can be characterized through Markov processes, aligning with the methodological insights presented in~\cite{Cao2021}. This modeling choice is grounded in the state-of-the-art of B-RAN approaches, thereby establishing a rigorous analytical framework. Specifically, the state of the system at any given moment, $t$, can be succinctly explicated through the mathematical expectation operator $E[i,j]$, where $i$ represents the pending requests awaiting aggregation into a block, and $j$ corresponds to the requests awaiting to be serviced. The incorporation of these variables underscores the nuanced interplay between the queue dynamics, reflecting the essential stages of pending requests en route to block inclusion and those poised for immediate service.

\begin{figure}
    \centering
    \includegraphics[width=0.9\linewidth]{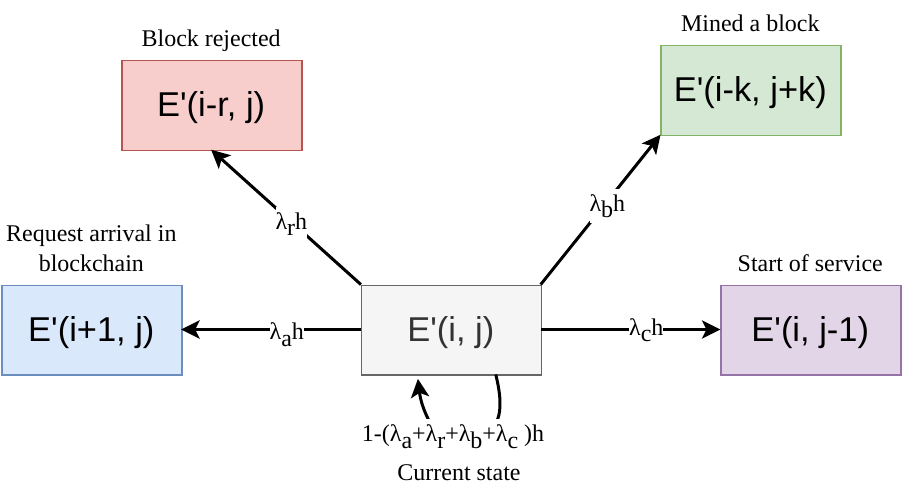}
    \caption{Markov Chain-based model of B-RAN.}
    \label{Fig:MC_model}
\end{figure}

As illustrated in~Fig.~\ref{Fig:MC_model}, the MC model is characterized by its current state, denoted as $E[i,j]$ at time $t$. It encompasses five distinct states, each representing specific configurations in the system's dynamics. The transition from one state to another occurs within a minimal time interval, denoted as $t+h$, where $h$ approaches zero. In more detail, the states of the MC model are detailed as follows.
\begin{itemize}
    \item $E'[i+1, j]$ denotes that a new request is received. This transition signifies an increment in the count of pending requests for blockchain aggregation ($i$), as one more request joins the existing set. Notably, the count of requests awaiting immediate service ($j$) remains unchanged, reflecting the constraint that only one procedure can occur at any given moment. This event occurs with a probability denoted as $p_a = \lambda_a \cdot h$.
    \item $E'[i-k, j+k]$ represents the scenario where a block is being mined. The probability of transitioning to this state is given by $p_b = \lambda_b \cdot h$ and depends upon the variable $k$, which represents the maximum number of requests that can be included in a single block. This transition can be categorized into two distinct cases depending on the number of pending requests and the block size. Specifically, if the number of pending requests, $i$, is less than or equal to the threshold $k$, then all pending requests are successfully mined, leading to an increase in the $j$ queue. In this case, the next state is represented as $E'[0, j+k]$. On the contrary, if the number of pending requests exceeds the threshold $k$, only a maximum of $k$ requests can be mined, while the rest of the requests remain pending. The mined requests contribute to an increase in the $j$ queue.
    \item $E'[i, j-1]$ expresses that a request is serviced and is characterized with a probability $p_c = \lambda_c \cdot h$. This transition signifies a decrement of 1 in the $j$ queue, reflecting the initiation of service for the respective block. Importantly, the count of pending requests, $i$, remains unchanged, as the commencement of service pertains solely to the queue of blocks and does not impact the queue of pending requests.
    \item $E'[i-r, j]$ denotes that a request is rejected due to various factors such as authentication issues or resource scarcity. In this case, $r$ expresses the number of rejected requests. This transition is governed by a rejection probability $p_r = \lambda_r \cdot h$. Consequently, it leads to a reduction of $r$ in the count of pending requests, $i$, as the block containing the rejected request is disposed of. Simultaneously, the $j$ queue, reflecting blocks awaiting service, remains unaffected, underscoring that the rejected block did not progress to the stage of mining.
    \item $E'[i, j]$ is the idle state. The probability of the system to remain in the same state, $p_i$, equals $1$ minus the sum of all the remaining rates, multiplied by $h$. This idle state implies that, at a given moment, the system remains in its current state without transitioning to any other state. The probability $p_i$ encompasses the combined likelihood of no new request arrivals, no rejected requests, no successful mining events and no completed service operations.
\end{itemize}

\section{Performance evaluation}\label{S:performance}
This section focuses on developing the analytical framework for evaluating the performance of the MC-based modelling of B-RAN. Specifically, Section~\ref{Ss:latency} provides information on the latency that can be achieved by the system, whilst Section~\ref{Ss:security} illustrates the security performance of the system.

\begin{figure*}
    \centering
    \includegraphics[width=0.9\linewidth]{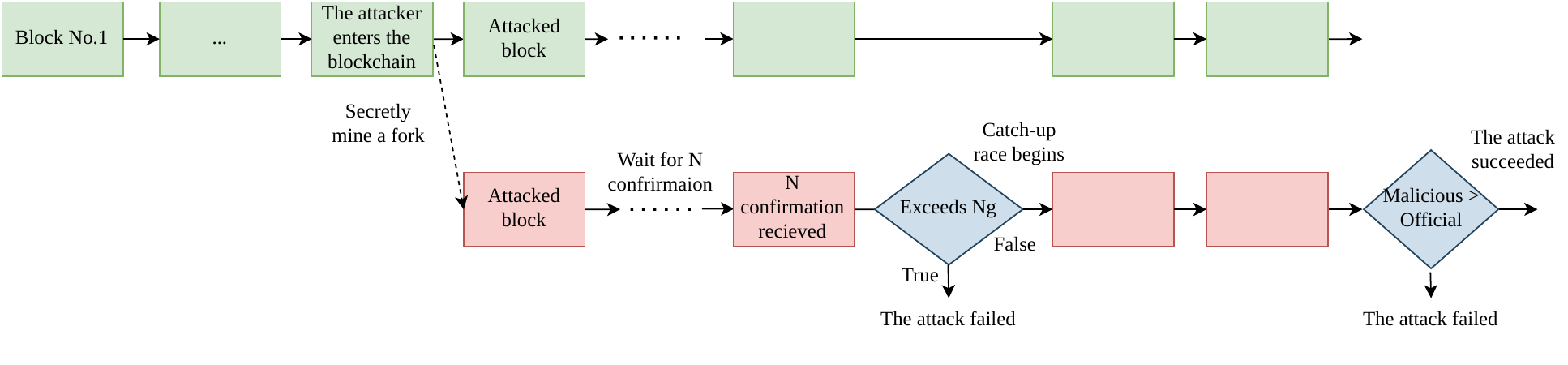}
    \caption{Procedural illustration of an alternate history attack in B-RAN.}
    \label{Fig:attack_diagram}
\end{figure*}

\subsection{Latency}\label{Ss:latency}
The incorporation of two queues within the proposed framework provides a structured approach to managing incoming requests and their subsequent processing in blockchain blocks, while at the same time modelling the temporal nature of the system. As stated above, the first M/M/1 queue handles the requests waiting to be included in blockchain, while and the M/M/s queue models the latency introduced by service initiation and processing. The two queues collectively contribute towards the end-to-end delay within the system. 

In the proposed model of B-RAN, the average waiting time within an M/M/1 queue can be analytically expressed as in~\cite{Cooper1981}
\begin{align}
    \tau_1 = \frac{1}{\lambda_b - \lambda_a} ,
\end{align}
where $\lambda_b$ represents the service rate and $\lambda_a$ denotes the arrival rate. In parallel, the latency induced by an M/M/s queue can be evaluated as in~\cite{Kelton1985}
\begin{align}
    \tau_2 = \frac{C(s, \frac{\lambda_a}{\lambda_c})}{s\lambda_c - \lambda_a} + \frac{1}{\lambda_c} ,
\end{align}
with the nominator of the first term expressing the Erlang C formula, which is dependent on parameters such as the number of simultaneously served users, $s$, the arrival rate, $\lambda_a$, and the service rate, $\lambda_c$. Furthermore, the final source of latency in the confirmation process that is considered in the present contribution, in which the average latency can be computed as
\begin{align}
    \tau_3 = \frac{N-1}{\lambda^b} ,
\end{align}
with $N$ representing the number of confirmations and $\lambda^b$ symbolizing the block generation rate. 

At this point, we utilize Little's Law in order to correlate the average latency with the length of the queue. According to Little's Law, the average number of transactions in a stable system is equal to the arrival rate multiplied by the average duration it spends in the system. Thus, the expected sojourn time can be expressed as 
\begin{align}
    \tau_s = \tau_1 + \tau_2 + \tau_3 ,
\end{align}
and is a quantitative measure of the duration that each service request remains in the specific states of the system. In other words, the $\tau_s$ is equal to the combined duration of waiting time and service time. Therefore, the average latency, $\tau_t$, of B-RAN is given by
\begin{align}
    \tau_t = \tau_s -\frac{1}{\lambda^c} .
\end{align}

Based on the aforementioned, the latency of the the B-RAN model can be bound in two respects. Specifically, the upper bound, capturing the maximum latency the model may incur, is composed of distinct components. Firstly, it involves an M/M/1 queue generated by transactions awaiting inclusion in a block. Additionally, an M/M/s queue accounts for transactions confirmed by the blockchain yet pending servicing. Furthermore, an additional latency factor arises from the requisite N number of confirmations necessary for each block. The expression of the upper bound can be expressed as 
\begin{align}
    \mathrm{L}_u = \frac{1}{\lambda^b-\lambda^a}+\frac{C\left(s, \frac{\lambda^a}{\lambda^c}\right)}{s \lambda^c-\lambda^a}+\frac{N-1}{\lambda^b} .
\end{align}

On the other hand, the lower bound expresses the minimum achievable latency within the model. In this scenario, the temporal aspect is governed by both the duration for a transaction to enter a block and the required number of confirmations. The expression for the lower bound is given by 
\begin{align}
    \mathrm{L}_l = \frac{1}{\lambda^b}+\frac{N-1}{\lambda^b}=\frac{N}{\lambda^b} .
\end{align}
All in all, this concise overview of the timing aspects of the considered model, underscores the pivotal role of queue selection in shaping the overall latency characteristics of the B-RAN framework.

\begin{figure*}
    \begin{align}
        S(N, \beta)= \begin{cases}1-\sum_{n=0}^N\left(\begin{array}{c}
        n+N-1 \\
        n
        \end{array}\right)\left(\frac{1}{1+\beta}\right)^N\left(\frac{\beta}{1+\beta}\right)^n\left(1-\beta^{N-n+1}\right) & \text { if } \beta<1 \\
        1 & \text { if } \beta \geq 1\end{cases}
        \label{Eq:P_attack}
    \end{align}
\end{figure*}
\subsection{Security}\label{Ss:security}
When it comes to secure and private communications, integrating blockchain technology into RAN shows promise in enhancing security and mitigating cyber threats. The decentralized and transparent nature of blockchain provides increased reliability against attacks that could compromise the integrity of RAN. However, there are still security concerns regarding blockchain that may create vulnerabilities in the system. One intriguing threat is the alternative history attack, where malicious entities attempt to manipulate transaction records within the blockchain network's history. This discussion explores the particularities of this attack and attempts to encapsulate how it can impact the procedures of B-RAN ultimately affecting its security and reliability.

In the scenario of alternative history attack, which is shown in Fig.~\ref{Fig:attack_diagram}, the attacker joins the blockchain during a regular event. After the official block is mined, the attacker secretly starts mining an altered fork; thus, creating a malicious version of the blockchain. The attacker's mining power, $\lambda_{m}$, is determined by the percentage of the attackers computing power relates to the blockchains mining rate, $\beta$. Despite differences in mining speeds between the two chains, other activities like request arrivals and service operations remain unchanged. Once the attacked block receives $N$ confirmations, the attacker initiates a race to catch up through mining. The attacker assesses how longer or shorter their malicious chain is compared to the chain. If this difference is below a threshold, $N_g$, then the attacker continues mining until the malicious chain surpasses the length of the official. On the other hand, if this difference exceeds $N_g$, the attacker stop mining; thus, considering the attack unsuccessful. On the contrary, when the malicious chain becomes longer than the original, the attack is considered successful.

The probability of a successful alternative history attack can be influenced by the attacker's relative mining rate, $\beta$, the needed number of confirmations, $N$, and the attacker's strategy, $N_g$. The expression of the successful attack probability is provided in~\eqref{Eq:P_attack} in the beginning of the next page~\cite{Ling2021}. Successfully mitigating the risk of intrusions requires a grasp and effective management of the aforementioned variables. 

\section{Numerical results}\label{S:results}
The purpose of this section is to evaluate the feasibility and effectiveness of the MC-based modeling of B-RAN that has been presented in this work. The numerical results provided emphasize the analytical discoveries and viewpoints about several pertinent scenarios and design issues. The validity of the analytical findings presented in this paper has been extensively verified by Monte Carlo simulations. Finally, this section is divided in two aspects of B-RAN, namely, timing performance and attack survivability.

\subsection{Timing performance}\label{Ss:timing}
\begin{figure}
    \centering
    \includegraphics[width=0.9\linewidth]{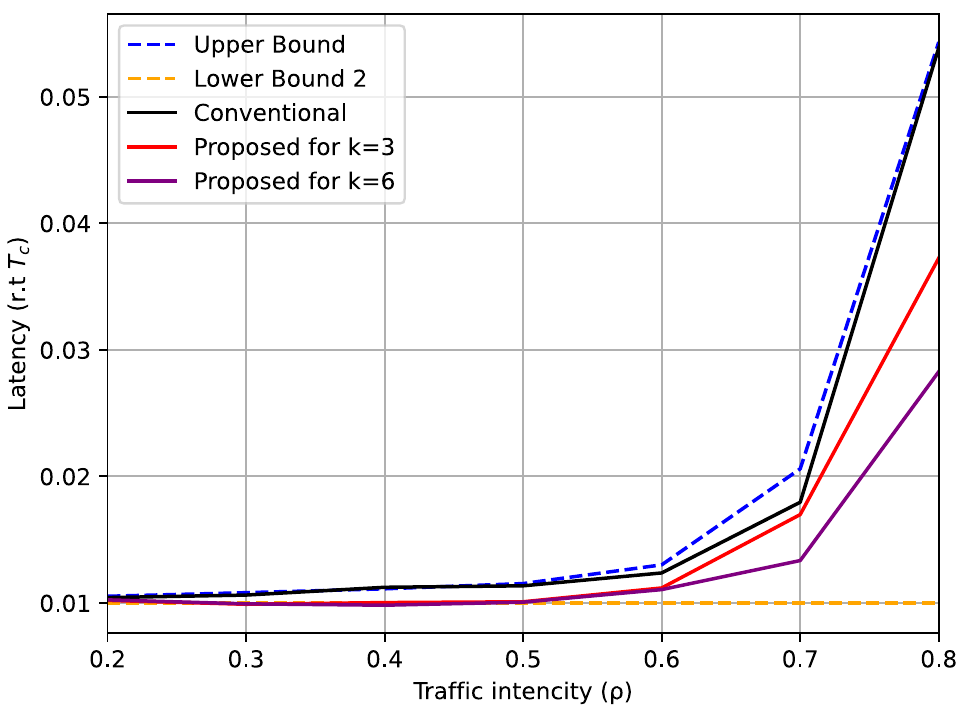}
    \caption{Average latency as a function of the traffic intensity for the proposed framework.}
    \label{Fig:Latency2}
\end{figure}
In Fig.~\ref{Fig:Latency2}, the average latency of the proposed model and its conventional counterpart is depicted as a function of the traffic intensity for different block sizes. Dashed lines mark the upper and lower boundaries of the average latency, while three cases are illustrated, namely, the conventional model and the proposed model with two distinct $k$ values. All plotted lines converge in low traffic scenarios, which indicates that, for modest levels of traffic, the two proposed models exhibit comparable latency profiles. As anticipated, when the traffic intensity increases, a corresponding rise in latency occurs across all three models. The conventional model diverges with an upward trajectory, resulting in a noticeable spike in latency. On the contrary, the proposed models, even as traffic intensity surges, display a smaller latency increase. Notably, as traffic attains maximum values, the conventional model manifests the highest latency, while the proposed model with a higher block size exhibits the lowest latency. This behaviour illustrates the superior performance of the proposed framework compared to the conventional model in scenarios of increased traffic intensity.

\begin{figure}
    \centering
    \includegraphics[width=0.9\linewidth]{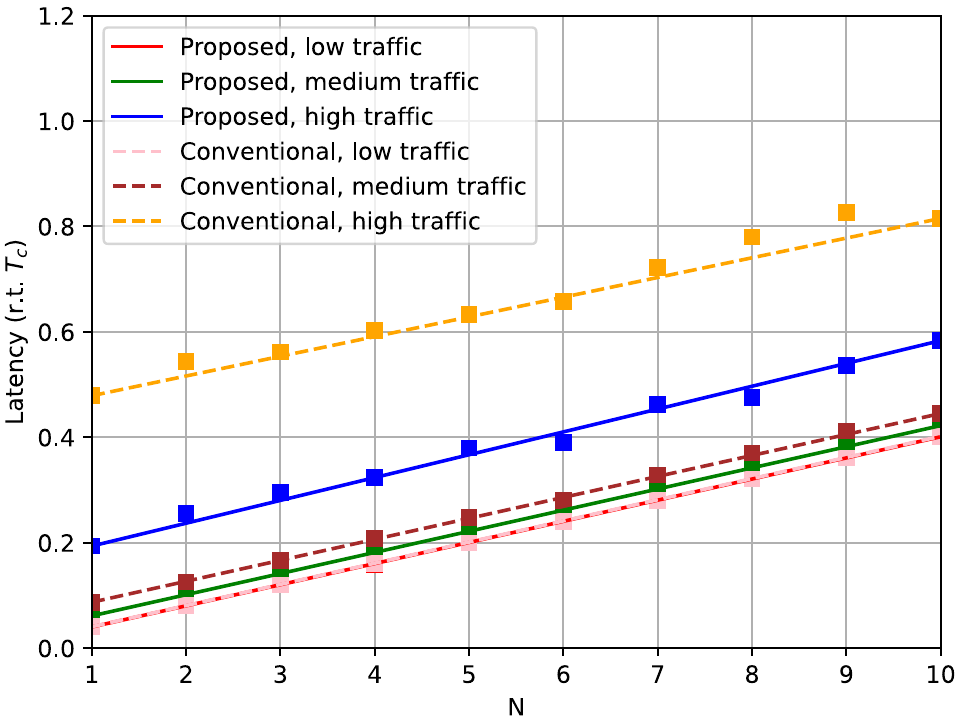}
    \caption{Average latency with regard to the number of confirmations for different traffic intensity scenarios.}
    \label{Fig:Latency1}
\end{figure}

Fig.~\ref{Fig:Latency1} illustrates a comparison of the average latency of the proposed framework and its conventional counterpart as a function of the number of confirmation under different traffic intensity conditions. It highlights how latency varies with different $N$ and $\rho$ values, offering an explanation of the impact of these variable on the system's performance. In the figure, three solid lines represent the proposed model, while three dashed lines display the conventional model. Square markers mark experimental results for both models in each scenario. Examining the trajectories of all scenarios reveals a shared pattern. Specifically, for a low number of $N$ confirmations, both models exhibit relatively low latency, which indicates higher timing efficiency. However, as the number of confirmations escalate, latency concurrently increases for all curves. Upon closer inspection, notable distinctions emerge between the proposed and conventional models. Particularly in scenarios characterized by low traffic intensities, it is observed that the two models exhibit similar latency across the entire range of $N$ values. However, for scenarios characterized by high traffic intensities, the conventional model consistently exhibits higher latency compared to its proposed counterpart. This trend illustrates that, under high traffic conditions, the proposed model consistently achieves lower latency compared to the conventional model. All in all, this analysis underscores the efficacy of the proposed model in optimizing latency performance, especially under conditions of elevated traffic intensity. 

\subsection{Attack survivability}\label{Ss:attack}

\begin{figure}
    \centering
    \includegraphics[width=0.9\linewidth]{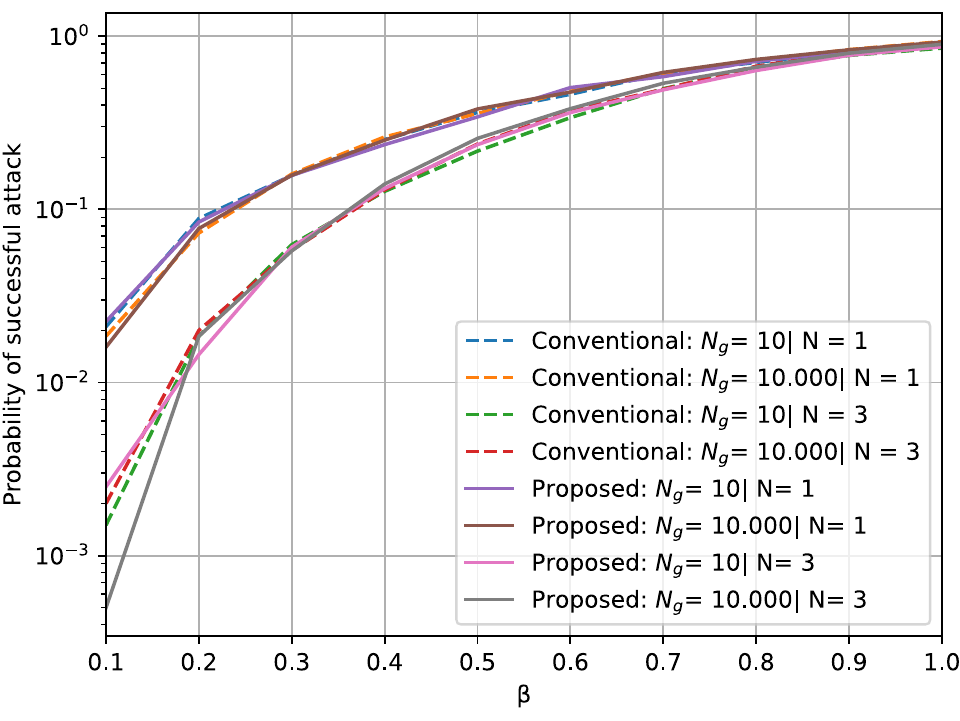}
    \caption{The likelihood of a successful alternate history attack varies depending on the attacker's computational resources, block confirmations, and chosen attack tactics.}
    \label{Fig:attack}
\end{figure}

In Fig.~\ref{Fig:attack}, the probability of successful attack is plotted against the attackers mining power for distinct cases of attack strategies and block confirmations. An variety of configurations are highlighted through 8 cases of both conventional and proposed models. Each subcase is characterized by varying $N_g$ and $N$ values. The figure reveals a convergence in the performance of both models. For $N = 1$ in both conventional and proposed models, they commence around $2 \times 10^{-2}$. Subsequently, with increasing $\beta$ values, there is a corresponding increase in the probability of successful attacks. Upon reaching $\beta=1$, the probability of a successful attack approaches $100\%$. Analogous patterns emerge for models with $N = 3$, initiating below $2 \times 10^{-3}$ and exhibiting an upward trajectory with higher $\beta$ values. Significantly, a convergence is observed for higher $\beta$ values, where both models attain a comparable profile. These findings suggest a consistent behavioral pattern for both models in diverse attack scenarios, unaffected by variations in the number of confirmations or $N_g$ values. Notably, the convergence at higher $\beta$ values underscores a robust convergence point, indicative of a shared vulnerability. The data implies that the efficacy of security protocols, as measured by successful attack probabilities, becomes increasingly pronounced at higher $\beta$ values, rendering both conventional and proposed models susceptible to comparable risk thresholds. 

\section{Conclusions}\label{S:conclusions}
In this paper we introduce a framework that aims to model the integration of blockchain technology in RAN; thus, overcoming the limitations of traditional models. By conducting simulations and comparing them with models our proposed framework not only expands its usability across various scenarios but also strikes a careful balance between reducing service latency and maintaining a strong security and privacy infrastructure. The results of these simulations consistently show that our framework outperforms conventional models. This reinforces the effectiveness of our approach in improving the efficiency and versatility of B-RAN. With the evolving landscape of wireless communication technologies there is an exciting opportunity to explore synergies with other cutting edge innovations, like edge computing, artificial intelligence and 5G networks.

\balance
\bibliographystyle{IEEEtran}
\bibliography{refs.bib}

\end{document}